\documentstyle[prl,aps,psfig]{revtex}
\newcommand{\BEQ}{\begin{equation}}     
\newcommand{\BEA}{\begin{eqnarray}}
\newcommand{\EEQ}{\end{equation}}       
\newcommand{\EEA}{\end{eqnarray}}
\def\be{\begin{equation}}
\def\ee{\end{equation}}
\def\bc{\begin{center}}
\def\ec{\end{center}}
\newcommand{\D}{{\rm d}}
\def\vecr{{\bf r}}
\def\qea{q_{\rm EA}}
\def\figun{
\begin{figure}[htb]
\centerline{\epsfxsize=3.25in\epsfbox{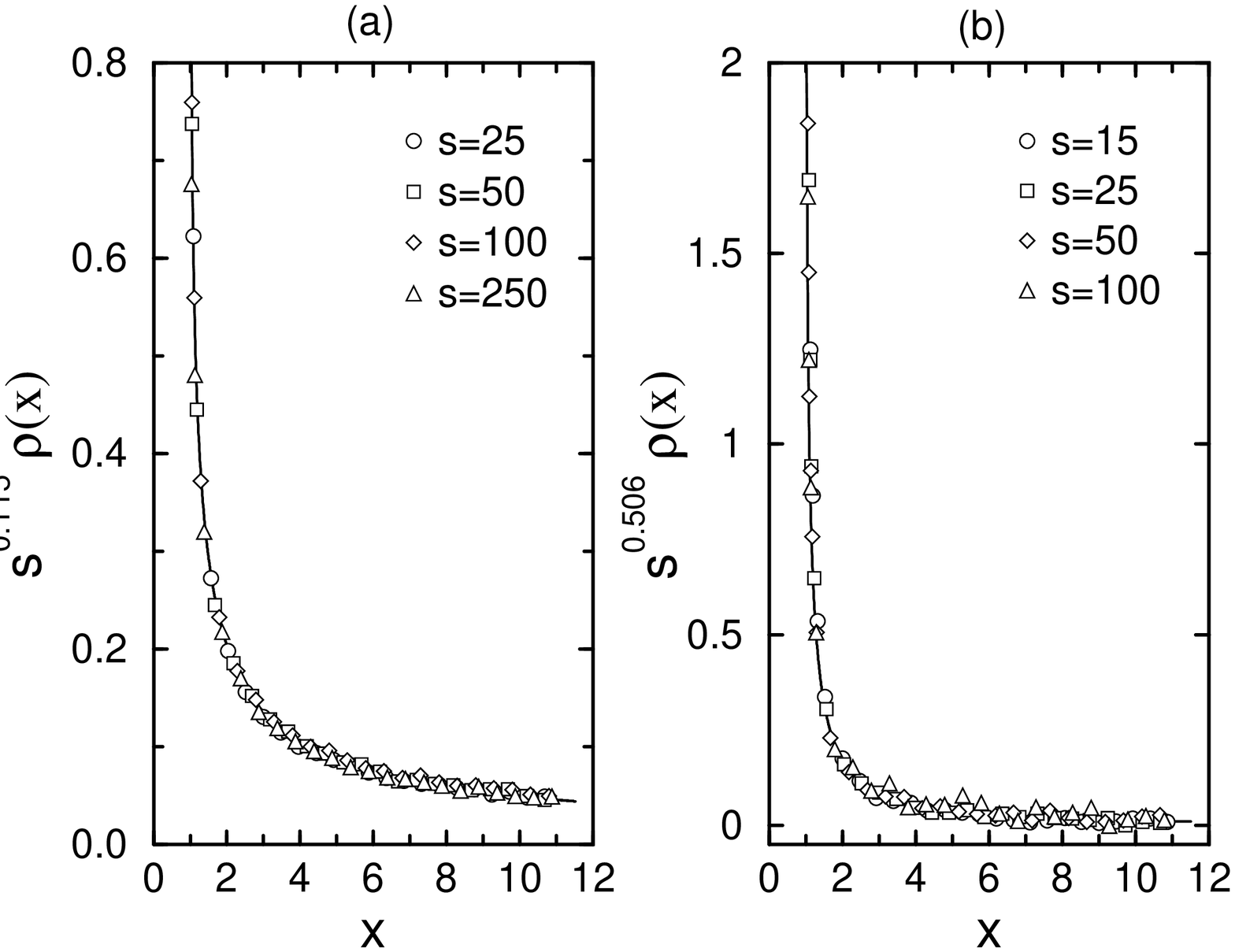}}
\caption{Scaling of the integrated response function $\rho$ for (a) the $2D$
and (b) the $3D$ Glauber-Ising model at criticality ($T=T_c$).
The symbols correspond to different waiting times.
The full curve is the conformal invariance prediction for $\rho$,
obtained by integrating~{\protect(\ref{R})}.
\label{Figur1}}
\end{figure}
}
\def\figde{
\begin{figure}[htb]
\centerline{\epsfxsize=3.25in\epsfbox{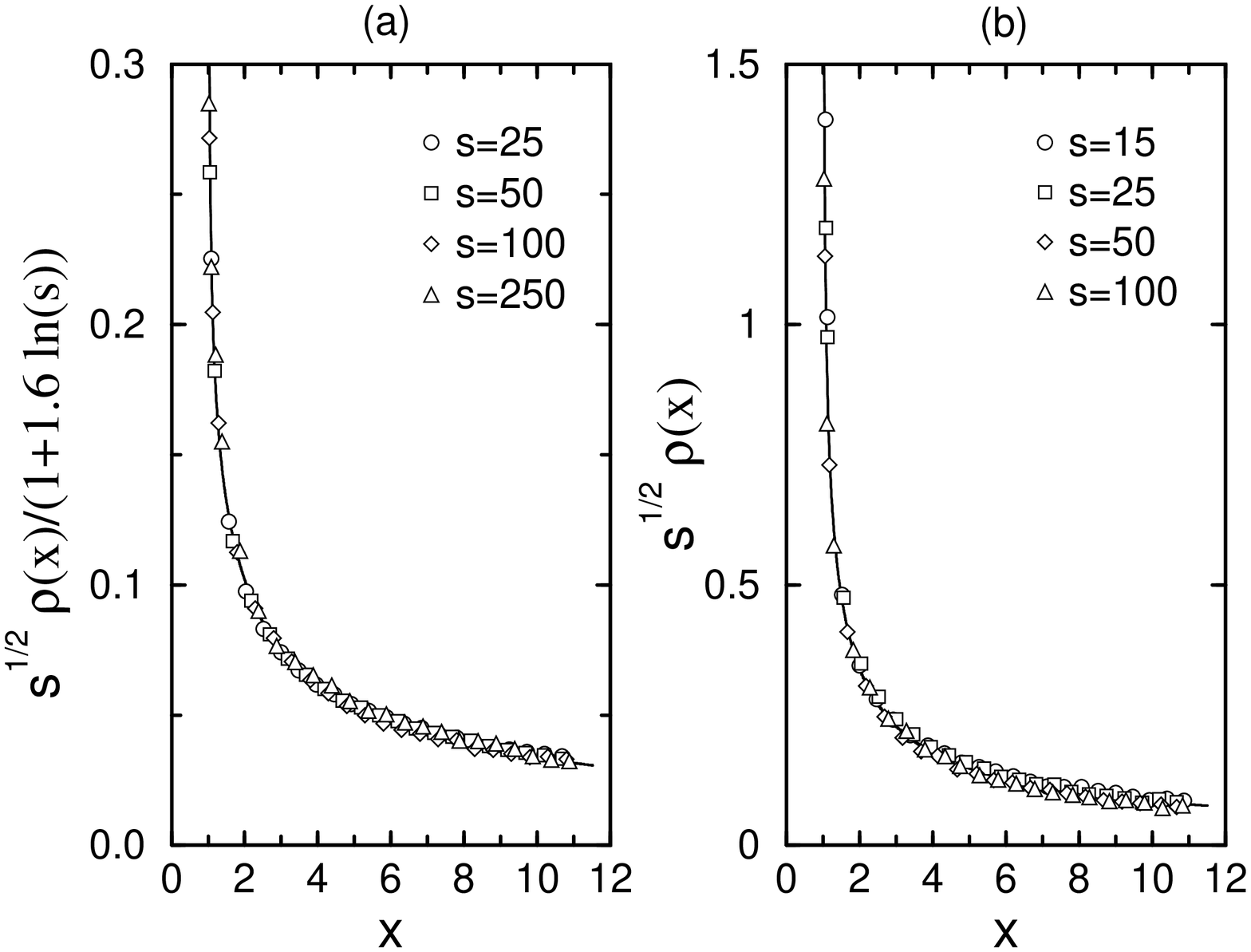}}
\caption{Scaling of the integrated response function $\rho$ for the
low-temperature Glauber-Ising model
(a) in $2D$ at $T=1.5$ and (b) in $3D$ at $T=3$.
The symbols correspond to different waiting times. The full curve is
obtained by integrating~{\protect(\ref{R})}.
\label{Figur2}}
\end{figure}
}
\def\tab{
\begin{table}[htb]
\caption{Critical temperature and exponents of the $2D$ and $3D$
Glauber-Ising model,
both in the phase-ordering regime ($T<T_c$)
and for critical dynamics ($T=T_c$).}
\label{Tabelle1}
\begin{tabular}{|l|l|l|l|}
\multicolumn{2}{|l|}{}       & $2D$       & $3D$   \\ \hline
\multicolumn{2}{|c|}{$T_c$}  & 2.2692     & 4.5115 \\ \hline
$z$         & $T=0$          & 2          & 2      \\
            & $T=T_c$        & 2.17       & 2.04   \\ \hline
$\lambda$   & $T=0$          & 1.25       & 1.50   \\
            & $T=T_c$        & 1.59       & 2.78   \\ \hline
$\beta/\nu$ & $T=T_c$        & $1/8$      & 0.517  \\
\end{tabular}
\end{table}
}

\begin{document}

\input epsf.sty
\twocolumn[\hsize\textwidth\columnwidth\hsize\csname %
@twocolumnfalse\endcsname

\draft

\widetext

\title{Aging, phase ordering and conformal invariance}

\author{Malte Henkel$^a$, Michel Pleimling,$^{a,b,*}$
Claude Godr\`eche,$^c$ and Jean-Marc Luck$^d$}

\address{
$^a$ Laboratoire de Physique des Mat\'eriaux,$^{**}$
Universit\'e Henri Poincar\'e Nancy I, B.P. 239,\\
F -- 54506 Vand{\oe}uvre l\`es Nancy Cedex, France\\
$^b$ Institut f\"ur Theoretische Physik I, Universit\"at Erlangen-N\"urnberg,
D -- 91058 Erlangen, Germany\\
$^c$ SPEC, CEA Saclay, F -- 91191 Gif-sur-Yvette Cedex, France\\
$^d$ SPhT,$^{***}$ CEA Saclay, F -- 91191 Gif-sur-Yvette Cedex, France}
\maketitle

\begin{abstract}
In a variety of systems which exhibit aging,
the two-time response function scales as $R(t,s)\approx s^{-1-a} f(t/s)$.
We argue that dynamical scaling can be extended towards conformal invariance,
obtaining thus the explicit form of the scaling function~$f$.
This quantitative prediction is confirmed in several spin systems,
both for $T<T_c$ (phase ordering)
and $T=T_c$ (non-equilibrium critical dynamics).
The $2D$ and $3D$ Ising models with Glauber dynamics are studied numerically,
while exact results are available for the spherical model
with a non-conserved order parameter, both for short-ranged and long-ranged
interactions, as well as for the mean-field spherical spin glass.
\end{abstract}
\phantom{.}
]

\narrowtext
Aging phenomena are observed in a broad variety of systems
with slow relaxation dynamics~\cite{Ang95}.
Aging behaviour is known to be more fully revealed by two-time quantities,
rather than by one-time quantities (see~\cite{Vin97,Bou98} for reviews).
The most commonly studied two-time quantities are
the correlation function $C(t,s)=\langle\phi(t)\phi(s)\rangle$
and the response function $R(t,s)=\delta\langle\phi(t)\rangle/\delta h(s)$,
where $\phi$ is the order-parameter field at some point,
and~$h$ is the conjugate local magnetic field,
while~$s$ is the waiting time and~$t$ the observation time.

Consider for definiteness, instead of a genuine glassy system,
the situation of a ferromagnetic model,
evolving at fixed temperature~$T$ from a disordered initial state.
In the high-temperature phase ($T>T_c$),
the relaxation time is small, so that
the system relaxes rapidly to equilibrium,
where $C(\tau)$ and $R(\tau)$ are time-translation invariant
(they depend only on the difference $\tau=t-s$) and
obey the fluctuation-dissipation theorem: $TR(\tau)=-\D C(\tau)/\D\tau$.
In the low-temperature phase ($T<T_c$), where coarsening takes place,
both $C(t,s)$ and $R(t,s)$ depend non-trivially on the ratio $x=t/s$
in the self-similar regime of phase ordering~\cite{Bra94}.
Similar features are also observed in the late stages of critical dynamics
$(T=T_c)$~\cite{Jan89,God00a,Lip00,God00b}.
The distance from equilibrium of an aging system
is usually characterized by the fluctuation-dissipation ratio
$X(t,s)$~\cite{Cug94,Ckp94,Vin97,Bou98}, such that
$T\,R(t,s)=X(t,s)\,\partial C(t,s)/\partial s$.

In the scaling regime where~$s$ and $\tau=t-s$ are simultaneously much larger
than the microscopic time scale (set to unity), the scaling laws
\BEQ
C(t,s)\approx s^{-b} F(t/s),\quad
R(t,s)\approx s^{-1-a} f(t/s)
\label{gl:CR}
\EEQ
are found to hold for a broad range of models~\cite{Vin97,Bou98,Bra94}.
Moreover, for $x\gg1$, i.e., $1\ll s\ll t$,
both scaling functions usually fall off as
\BEQ
F(x)\sim f(x)\sim x^{-\lambda/z},
\label{gl:Skalf}
\EEQ
where~$z$ is the dynamic critical exponent
and $\lambda$ is the autocorrelation exponent~\cite{Hus89}.
For a ferromagnetic model, with scalar non-conserved order parameter,
(\ref{gl:CR}) holds at criticality ($T=T_c$)~\cite{Jan89,God00b},
with $a=b=2\beta/(\nu z)=(d-2+\eta)/z$,
where $\beta$, $\nu$, $\eta$ are static exponents,
while~$z$ is known from equilibrium critical dynamics.
In the phase-ordering regime ($T<T_c$),~(\ref{gl:CR}) only holds
in the aging regime, where $C(t,s)$ decays from its plateau value
$\qea=M_{\rm eq}^2$ to zero
($M_{\rm eq}$ is the spontaneous magnetization).
One has $b=0$ and $F(1)=\qea$.
There seems to be no general result for the response exponent:
$a=1/2$ for the Glauber-Ising model, both in one dimension~\cite{God00a,Lip00} 
and in higher dimensions~\cite{Ber99}, while $a=d/2-1$ for the spherical model
in dimension $d>2$~\cite{Cug95,Zkh00,God00b}. 
We stress that the specific scaling forms (\ref{gl:CR},\ref{gl:Skalf}) only
hold for a completely {\em disordered\/} initial state. If there are 
correlations in the initial state, similar but different scaling forms 
hold \cite{New90,Ber01}. 

At present, there is no general principle to predict the form of correlation 
and response functions in non-equilibrium systems.
On the other hand, for an {\em equilibrium} critical
point (where formally $z=1$), scale invariance can be extended to conformal
invariance~\cite{Car96,Fra97,Hen99}. Scale invariance
implies that correlators transform covariantly under dilatations, i.e.,
scale transformations which are spatially uniform. Conformal transformations
are {\em local} scale transformations with a position-dependent dilatation
factor $b=b(\vecr)$, but such that angles are conserved.
This is already enough to fix the form of an equilibrium correlator at
criticality in any space dimension. Furthermore, $2D$ conformal invariance
yields exact values for the entire set of critical exponents, the exact form
of all $n$-point correlators, a classification of the universality classes,
and much more~\cite{Bel84} (see~\cite{Fra97,Hen99} for reviews).

Is a similar extension of the scale invariance of~(\ref{gl:CR})
also available for non-equilibrium systems?
Indeed, we shall present evidence that this might be so.
First, we argue~\cite{Hen01} that
the two-time response function $R(t,s)$ should transform covariantly
under the action of conformal transformations in time.
This assumption is then shown to imply the scaling form
\BEA
R(t,s)&\approx&r_0\,(t/s)^{1+a-\lambda/z}(t-s)^{-1-a},\nonumber\\
\hbox{i.e.,}\;\;\;f(x)&=&r_0\,x^{1+a-\lambda/z}(x-1)^{-1-a}.
\label{R}
\EEA
The scaling function $f(x)$ is thus entirely fixed,
up to the normalization constant $r_0$,
by the two exponents~$a$ and $\lambda/z$ entering~(\ref{gl:CR})
and~(\ref{gl:Skalf}).
This explicit scaling form~(\ref{R}) of the response function
is the main result of this Letter.
It is expected to hold throughout the aging regime,
for any value of the ratio $x=t/s > 1$.
The power law~(\ref{gl:Skalf}) is recovered
for well-separated times ($x\gg 1$) \cite{Note1}.

The prediction~(\ref{R}) will be corroborated by numerical simulations in 
the $2D$ and $3D$ Ising model with Glau\-ber dynamics.
In addition, exact results for the spher\-ical
model with a non-conserved order parameter (including
spatially long-ranged interactions and/or quenched disorder) also
reproduce~(\ref{R})~\cite{Jan89,God00b,Ckp94,Cug95,New90,Hen94,Can01}.
This confirmation provides, for the first time, evidence for conformal
invariance in nonequilibrium and aging phenomena.

We now sketch the line of reasoning leading to~(\ref{R}).
The full calculation will be given elsewhere~\cite{Hen01}.
To begin, we ask: what
space-time symmetries are consistent with dynamical scale invariance
$t\to b^z t$, $\vecr\to b\vecr$, where~$z$ is the dynamical
exponent~\cite{Hen97,Hen01}?
A similar question has already been successfully raised for equilibrium
systems with strongly anisotropic scaling~\cite{Hen97,Ple01}. While
in equilibrium systems the correlation functions (of quasiprimary
operators~\cite{Bel84}) are expected to transform
in a simple way, for non-equilibrium systems it is rather the response
functions which will take this role, as argued long ago~\cite{Car85}.
We expect the requested extension of dynamical scaling to contain the M\"obius
transformations of time: $t\mapsto t'=(\alpha t+\beta)/(\gamma t+\delta)$, with
$\alpha\delta-\beta\gamma=1$, since these occur in the two known special cases,
namely conformal invariance for $z=1$ and Schr\"odinger invariance for $z=2$.
It turns out that this condition is already sufficient to fix the form of the
infinitesimal generators. In one space dimension, to which we restrict for
notational simplicity, it can be shown that for non-equilibrium systems
one may write~\cite{Hen01}
\BEA
X_{-1}&=&-\partial_t,\quad X_0=-t\partial_t-(1/z)r\partial_r,
\nonumber\\
X_{1}~~&=&-t^2\partial_t-(2/z)tr\partial_r-\beta r^2\partial_r^{2-z},
\label{gene}
\EEA
where $\beta$ is a constant related to ``mass''~\cite{Hen97,Hen01}.
The generators $X_n$ satisfy the commutation
relations $[X_n, X_m] = (n-m)X_{n+m}$ (with $n,m\in\{-1,0,1\}$) of the Lie
algebra of the conformal group.

Equation~(\ref{gene}) forms the basis for the derivation of~(\ref{R}).
Time translations are generated by $X_{-1}$.
In order to make the above construction applicable to non-equi\-lib\-rium
situations, we must discard the latter, and only require covariance under the
subalgebra $\cal S$ generated by $X_0$ and $X_1$ (see~\cite{Hen94}).
It is clear from the form of the generators
that the initial line $t=0$ is invariant under the action of $\cal S$.
Now, consider a general response function
$G=\langle\phi_1(t_1,r_1)\widetilde{\phi}(t_2,r_2)\rangle$,
where the field $\phi_1$ is characterized by its scaling
dimension $x_1$ and ``mass'' $\beta_1$, and the response field
$\widetilde{\phi}_2$ (see e.g.~\cite{Car96}) has scaling dimension $x_2$
and ``mass'' $\beta_2$. Then the covariance of~$G$ under the local scale
transformations in $\cal S$ is expressed by the conditions~\cite{Hen01}
$X_0 G = (\zeta_1 +\zeta_2) G$, $X_1 G =(2\zeta_1 t_1 + 2\zeta_2 t_2)G$,
where $\zeta_i = x_i/z$ ($i=1,2$). Moreover, we require spatial translation
invariance, thus $G=G(t_1,t_2; r_1-r_2)$. 
This is always satisfied if $\beta_1+(-1)^{2-z}\beta_2=0$. 
We can now set $r=r_1-r_2=0$ and obtain the response function $G=R(t_1, t_2)$.
Then the generators~(\ref{gene}) reduce to the standard conformal generators
(see e.g.~\cite{Fra97,Hen99,Bel84}),
and~$R$ satisfies the differential equations
\BEA
( t\partial_t + s\partial_s +\zeta_1 +\zeta_2) R(t,s) &=& 0,
\nonumber\\
( t^2\partial_t + s^2\partial_s + 2\zeta_1 t + 2\zeta_2 s ) R(t,s) &=& 0,
\label{gl:Xgen}
\EEA
hence $R(t,s) = r_0 (t/s)^{\zeta_2-\zeta_1} (t-s)^{-\zeta_1-\zeta_2}$.
An identification of exponents with~(\ref{gl:CR}),~(\ref{gl:Skalf}) 
yields~(\ref{R}).

The prediction~(\ref{R}) will now be checked against
results for various model systems.
We begin with a novel numerical investigation of the Ising model,
on square or cubic lattices, with periodic boundary conditions,
and Glauber or heat-bath dynamics.
Because the instantaneous response function $R(t,s)$
is too noisy to be measured in a simulation,
we consider instead the integrated response function~\cite{Bar98,God00b}
\BEQ
\rho(t,s) = T\int_{0}^{s}\!\D u\, R(t,u)\approx
\left( T/h\right) M_{\rm TRM}(t,s),
\EEQ
where $M_{\rm TRM}(t,s)$ is the thermoremanent magnetization, i.e.,
the magnetization of the system at observation time~$t$ obtained after applying
locally a small magnetic field~$h$ between the initial time
$t=0$ and the waiting time $t=s$.
This quantity can be readily measured, either in TRM experiments,
or in numerical simulations~\cite{Bar98}.
The data shown in Figures~\ref{Figur1} and~\ref{Figur2}
have been obtained in this way, and averaged
over at least 1000 different realizations of
systems with $300\times 300$ spins in $2D$ and $50\times 50\times 50$ spins
in $3D$. Larger systems were also simulated, in order to check
for finite-size effects.
Table~\ref{Tabelle1} contains the numerical values of exponents
used in the subsequent analysis.

Figure~\ref{Figur1} displays our results for $\rho(t,s)$
in the scaling regime at criticality
(data corresponding to $1\sim\tau\ll s$ have been discarded).
From~(\ref{gl:CR}) we expect a data collapse
if $s^a\rho(t,s)$ is plotted against $x=t/s$, and this is indeed the case.
Having thus confirmed the expected scaling,
we can compare with the prediction~(\ref{R}).
We find complete quantitative agreement, after adjusting only one parameter,
the overall normalization constant $r_0$.
Figure~\ref{Figur2} shows our results in $2D$ and $3D$,
in the scaling regime and for a fixed temperature below $T_c$.
As $a=1/2$~\cite{Ber99},
we expect that $s^{1/2}\rho(t,s)$ only depends on $t/s$.
This is indeed the case in $3D$, but for $2D$ the situation is more complicated.
Analytical calculations in the spirit of the OJK approximation~\cite{Bra94}
reveal the presence of extra logarithms:
$\rho(t,s)\approx s^{-1/2}\ln s\;f(t/s)$~\cite{Ber99}.
Logarithmic corrections to scaling are not so rare,
even in the realm of $2D$ equilibrium conformal theories\cite{Gur93,Hen99}.
We therefore propose the heuristic ansatz
\BEQ
\rho(t,s)\approx s^{-1/2}\left( r_0 + r_1\ln s\right) E(t/s),
\EEQ
where $r_0$, $r_1$ are non-universal constants,
and $E(x)$ is a scaling function. It is apparent from Figure~\ref{Figur2}(a)
that we thus obtain a satisfactory scaling.
We have checked that the same scaling form holds
in the entire low-temperature phase, where $r_0$, $r_1$ depend on~$T$.
We again find complete agreement between the form
of the scaling function $E(x)$ and the prediction~(\ref{R}).

We now turn to confirmations of~(\ref{R})
by means of available analytical results.
For the ferromagnetic spherical model~\cite{God00b},
which can alternatively be described in terms of a continuum field
theory~\cite{Jan89,New90},
the scaling expression of the response function has been derived
in any dimension $d>2$.
It reads $R(t,s)\approx(4\pi s)^{-d/2} f(x)$,
where the scaling function is,
in the ordered phase ($T<T_c$)~\cite{New90,God00b},
\BEQ
f(x) = x^{d/4} (x-1)^{-d/2},
\label{f2x}
\EEQ
and at the critical point ($T=T_c$)~\cite{Jan89,God00b},
\BEQ
f(x) =\left\{\begin{array}{ll}
x^{1-d/4} (x-1)^{-d/2} &\!\!~ (2<d<4),\\
(x-1)^{-d/2} &\!\!~ (4<d).
\end{array}\right.
\label{sphcrit}
\EEQ
These expressions are in full agreement with~(\ref{R}).
The second expression of~(\ref{sphcrit}),
corresponding to the mean-field situation,
coincides with the result for a free (Gaussian) field~\cite{Ckp94},
as could be expected.

The spherical model has the peculiarity that the dynamical
exponent $z=2$ throughout. In that case, the response functions are expected
to transform covariantly under the Schr\"odinger group~\cite{Hen94}.
The full space-time dependent response reads
\BEQ
\left\langle\phi(t,\vecr_1)\widetilde{\phi}(s,\vecr_2)\right\rangle=R(t,s)\,
\exp\!\left(\!-\frac{\cal M}{2}\frac{(\vecr_1-\vecr_2)^2}{t-s}\!\right),
\EEQ
with $R(t,s)$ given by~(\ref{R}), and where the ``mass'' $\cal M$ is a constant
(in~(\ref{gene}), $\beta={\cal M}/2$ for $z=2$).
A comparison with the exact spherical model results, both at and below
$T_c$~\cite{Jan89,New90},
also permits to confirm this fully~\cite{Hen94,Hen99}.

Recently, correlation and response functions have been calculated exactly
for the spherical model with long-range interactions of the form
$J(\vecr)\sim |\vecr|^{-d-\sigma}$~\cite{Can01}.
For $d>2$ and $\sigma>2$, the spherical model with short-ranged interactions, 
discussed above, is recovered.
On the other hand, for $d>2$ and $0<\sigma<2$, or $d\leq 2$ and $0<\sigma<d$,
the dynamical exponent reads $z=\sigma$ below criticality,
while the response function scales as~\cite{Can01}
\BEQ
R(t,s)\approx r_0\, (t/s)^{d/(2\sigma)} (t-s)^{-d/\sigma},
\EEQ
which again agrees with~(\ref{R}).
This example illustrates that spatially long-ranged interactions need not
destroy conformal invariance in non-equilibrium situations,
in contrast to the situation of conformal invariance {\em at} equilibrium.
Moreover, for the mean-field spherical spin glass~\cite{Cug95},
the response function reads
$R(t,s)\sim (t/s)^{3/4}(t-s)^{-3/2}$ in the low-temperature aging regime.
This result also agrees with~(\ref{R}).
It coincides with~(\ref{f2x}) for $d=3$,
as a consequence of the known similarity between the $3D$ spherical ferromagnet
and mean-field spin glass~\cite{Zkh00}.
Finally, note that for the simple random walk $R(t,s)=r_0=\mbox{\rm cste.}$ for 
$t>s$, see \cite{Ckp94}, which is consistent with (\ref{R}) with
exponents $a=-1$ and $\lambda/z=0$. 

In conclusion, the dynamical scale invariance realized in non-equilibrium
aging phenomena apparently generalizes towards (a subgroup of)
conformal invariance. As a first consequence, we obtained the explicit
scaling expression~(\ref{R}) for the two-time response function $R(t,s)$,
whose functional form only depends on the values of the exponents
$a$ and $\lambda/z$. Our prediction~(\ref{R})
has been checked against analytical and numerical results
for several spin systems with a non-conserved order parameter. 
The entire evidence available at present comes from classical systems with a
disordered initial state. Different initial conditions may lead to a modified
scaling behaviour \cite{New90,Ber01} and the applicability of conformal
invariance to these remains to be studied. 
The problem of identifying the full set of physical conditions on the systems 
which obey~(\ref{R}) remains open. 
However, validity of (\ref{R}) might extend to a broader class of systems than 
studied here, possibly including some realistic glassy systems. 
Finally, it appears that the conditions for the applicability
of conformal invariance in non-equilibrium situations
(such as the value of~$z$, or the presence of long-ranged interactions)
are {\em less} restrictive than for conformal invariance
at equilibrium critical points. 
We hope that the ideas presented might shed some light on some of the
standing questions of aging phenomena \cite{Note2}. 

MH thanks the SPhT Saclay, where this work was started, for warm hospitality.
We thank the CINES Montpellier for providing substantial computer time
(projet pmn2095).


\tab
\figun\figde
\end{document}